\documentclass[11pt,fleqn]{article}
\usepackage{amsfonts,amsmath,amssymb,amscd,graphicx,microtype,hyperref}

\usepackage{amsfonts,amssymb,cite}
\usepackage{epsf,graphicx}

\def\noi{\noindent}

\newcommand{\Title}[1]{\noi {{\Large\bf #1}}\\[1ex]}

\def\Aunames#1{\noi{\bf #1}}
\def\auth#1{${}^{#1}$}
\def\Addresses#1{\medskip\noi \protect
	\begin{description}\itemsep -3pt {\it #1} \end{description}}
\def\addr#1#2{\item[${}^{#1}$]{\it #2}}

\newcommand{\Abstract}[1]{\vskip 2mm \begin{center}
        \parbox{16.4cm}{\small\noi #1} \end{center}\medskip}

\def\email#1#2{\footnotetext[#1]{e-mail: #2}\addtocounter{footnote}{1}}

\def\nqq{\hspace*{-2em}}

\def\Jl#1#2{#1 {\bf #2},\ }

\def\ApJ#1 {\Jl{Astroph. J.}{#1}}
\def\CQG#1 {\Jl{Class. Quantum Grav.}{#1}}
\def\DAN#1 {\Jl{Dokl. AN SSSR}{#1}}
\def\GC#1 {\Jl{Grav. Cosmol.}{#1}}
\def\GRG#1 {\Jl{Gen. Rel. Grav.}{#1}}
\def\JETF#1 {\Jl{Zh. Eksp. Teor. Fiz.}{#1}}
\def\JETP#1 {\Jl{Sov. Phys. JETP}{#1}}
\def\JHEP#1 {\Jl{JHEP}{#1}}
\def\JMP#1 {\Jl{J. Math. Phys.}{#1}}
\def\NPB#1 {\Jl{Nucl. Phys. B}{#1}}
\def\NP#1 {\Jl{Nucl. Phys.}{#1}}
\def\PLA#1 {\Jl{Phys. Lett. A}{#1}}
\def\PLB#1 {\Jl{Phys. Lett. B}{#1}}
\def\PRD#1 {\Jl{Phys. Rev. D}{#1}}
\def\PRL#1 {\Jl{Phys. Rev. Lett.}{#1}}

\def\lal{&&\nqq {}}

\def\beq{\begin{equation}}
\def\eeq{\end{equation}}
\def\bear{\begin{eqnarray}}
\def\bearr{\begin{eqnarray} \lal}
\def\ear{\end{eqnarray}}
\def\earn{\nonumber \end{eqnarray}}

\newcommand{\be}{\begin{equation}\label}
\newcommand{\ee}{\end{equation}}
\newcommand{\prt}{\partial}
\newcommand{\p}{\prime}
\newcommand{\wh}{\widehat}

\def\B{{\mathbb B}}
\def\C{{\mathbb C}}

\begin{document}

\Title{Algebrodynamics: Shear-Free Null Congruences and New Types of Electromagnetic Fields}
\Aunames{Vladimir~V.~Kassandrov\auth{a,1}, Joseph~A.~Rizcallah \auth{b,2} and Ivan~A.~Matveev \auth{c,3}}

\Addresses{
\addr a {Institute of Gravitation and Cosmology, Peoples' Friendship University of Russia,  Miklukho-Maklaya str., 6, Moscow, 117198, Russia}
\addr b {School of Education, Lebanese University, Beirut P.O. Box 6573/14, Lebanon}
\addr c {Federal Research Center ``Computer Science and Control'' of the Russian Academy of Sciences, Vavilov Str., 44/2, Moscow, 119333, Russia}	
	}

\Abstract
  {We briefly present our version of noncommutative analysis over matrix algebras, the algebra of biquaternions ($\mathbb B$) in particular. 
We demonstrate that any $\mathbb B$-differentiable function gives rise to a null shear-free congruence (NSFC) on the $\mathbb B$-vector space $\mathbb{C}\bf M$ and on its Minkowski subspace $\bf M$. 
Making use of the Kerr--Penrose correspondence between NSFC and twistor functions, we obtain the general solution to the equations of $\mathbb B$-differentiability and demonstrate that the source of an NSFC is, generically, a world sheet of a string in $\mathbb {C}\bf M$. 
Any singular point, caustic of an NSFC, is located on the complex null cone of a point on the generating string. Further we describe symmetries and associated gauge and spinor fields, with two electromagnetic types among them. 
A number of familiar and novel examples of NSFC and their singular loci are described. 
Finally, we describe a conservative algebraic dynamics of a set of identical particles on the ``Unique Worldline'' and discuss the connections of the theory with the Feynman--Wheeler concept of ``One-Electron Universe''.}


\email 1 {vkassan@sci.pfu.edu.ru}
\email 2 {joeriz68@gmail.com}
\email 3 {matveev@frccsc.ru}

\section{Introduction}
Congruences of light rays, with zero shear in particular, play a crucial role in general relativity and twistor aspects of geometry of the flat Minkowski space--time 
$\bf M$ (or its Kerr--Schild curved generalization). 
They relate also to the conditions of differentiability for the biquaternion-valued functions, which may be put into correspondence with fundamental physical fields (in the framework of the so-called {\it algebrodynamical} approach, see below). 

The general solution to the defining equations of null shear-free (geodesic) congruences (NSFC) is given by the {\it Kerr theorem} and can be formulated in the following implicit algebraic form~\cite{Penrose1}:
\be{Kerr}
 \Pi (\xi, iX\xi) = 0 \;
 \ee
where $\Pi(\xi,\tau),~\tau:=iX\xi$ is an {\it arbitrary homogeneous} (and almost everywhere analytical) function of three {\it projective} components of the twistor $\{\xi,\tau\}$ on $\bf M$, i.e., a pair of two-spinors linked at every space--time point $X=X^+=\{X^{AA^\p}\},~A=A^\p= 0,1$, via the Penrose's 
{\it incidence relation} {(see, e.g.,~\cite{Penrose1}, ch. 6)}, 
 
\be{inc}
 \tau=iX\xi~~~(\tau^A=iX^{AA^\p}\xi_{A^\p}).
\ee

\textls[-15]{At a fixed point $X \in \bf M$, resolving \eqref{Kerr} w.r.t. the ratio of components of the spinor $\xi_{A^\p}=\xi_{A^\p} (X)$, one arrives at a null four-vector field $k_\mu = \xi_A \xi_{A^\p},~\mu= 0,1,2,3$, tangent to the lines of the congruence, which is necessarily shear-free. 
According to the Kerr--Penrose theorem, all (analytical) NSFC can be constructed via the algebraic procedure presented above.} 

{\it Caustics}-singularities of NSFC are defined by the condition 
\be{caust}
 \Pi^\p (\xi, X\xi) = 0 \;,
\ee 
i.e., correspond to the space--time points at which the {\it total derivative} of $\Pi$ w.r.t. the {\it ratio} of spinor components turns to zero. 
Obtaining the latter from \eqref{caust} and substituting the result into \eqref{Kerr}, one arrives at the {\it equation of motion} 
\be{eqmotion}
 S(X):=\Pi(\xi(X), iX\xi(X)) = 0, 
\ee
which defines the shape of the singular locus of the congruence (at a fixed moment 
of time) together with its evolution in time. 
It is easy to check (see, e.g.,~\cite{Burin2}) that any such function $S(X)$ satisfies the {\it complex eikonal equation} (CEE). 
Moreover, the ratio of spinor components satisfies both the CEE and wave d'Alembert equation~\cite{Wilson, Visser2007}.
 
In the framework of the so-called {\it algebrodynamics} (see, e.g.,~\cite{AD3, AD4} and references therein), such singularities (when bounded in three-space) are identified with {\it particle-like} formations. 
A remarkable example of these in general relativity is represented by the Kerr {\it singular ring}-caustic of the twofold static and axisymmetric Kerr congruence. 
It possesses some of the quantum numbers of an elementary fermion (Dirac electron)~\cite{Carter, Burin, Burinskii2014}. 
\mbox{Newman et al.~\cite{Lind,Newman1}} have shown that the Kerr-like congruences may be understood as 
generated by a ``virtual'' point charge moving along a complex world line $Z_\mu = Z_\mu (\sigma),~\sigma \in \mathbb{C}$ in the {\it complex extension} $\mathbb{C}\bf M$ of the Minkowski space--time $\bf M$ and emitting therein rectilinear complex ``light-like rays'' (that is, null complex straight lines)~\cite{Cramer} (p. 382). 
The restriction of such a complex congruence onto the real $\bf M$ gives rise to an NSFC, in particular to the Kerr one (in the case of a virtual charge at rest in $\mathbb{C}\bf M$). 

In Section \ref{sec2}, we briefly present an original approach to the construction of noncommutative analysis, primarily over the algebra of biquaternions ($\mathbb B$). 
In Section \ref{sec3}, we obtain the general solution of the corresponding $\mathbb B$-differentiability conditions and establish their equivalence with twistor structures and the Kerr--Penrose condition defining any NSFC.

However, the above construction covers only a narrow subclass of NSFC. 
Below, in Section \ref{sec3}, we demonstrate that the source of an NSFC of a generic type is a {\it complex string}. 
It turns out that any singular point of an NSFC belongs to a complex null cone of a point on the string.    

Section \ref{sec4} is devoted to symmetries and gauge and spinor fields associated with the $\mathbb B$-differentiability equations (precisely, with NSFC). 
In particular, we introduce two types of electromagnetic fields that satisfy the free Maxwell equations on their solutions. 
In Section \ref{sec5}, we exhibit some familiar and novel examples of NSFC and its singularities. 
 
In Section \ref{sec6}, we briefly review our papers~\cite{Ildus1,Ildus3} in which, following the ideas of Wheeler and Feynman on the so-called {\it Unique Worldline} and the algebrodynamical approach, a conservative algebraic dynamics of a set of identical point particles has been elaborated. Finally, in Section \ref{sec7}, the relativistic version of this construction is presented.  

\section{$\B$-Differentiability and Algebrodynamics}\label{sec2}
  
The calculus of functions of quaternionic ($\mathbb Q$) variable, the analog of complex ($\mathbb C$) analysis, has a long history and in the opinion of, say, M. Atiah~\cite{Atiyah} (p. 184) or R. Penrose~\cite{PenroseBook} (p. 201) is either not yet solved or a principally insoluble problem. 
The most known attempt to build a $\mathbb Q$-analogue of $\mathbb C$-analysis belongs to R. Fueter~\cite{Fueter,Fueter2}, while in~\cite{Gursey, Gursey1}, the authors tried to use the corresponding holomorphic conditions in the framework of chiral and gauge models. 
Other attempts have also been undertaken, in particular, by A. Deavours~\cite{Deav}, A. Yu. Khrennikov~\cite{Khren}, Schwartz~\cite{Schwartz2009} and others. 

However, Fueter's and other approaches are in many aspects unsatisfactory. 
Specifically, if one requires, in analogy with the $\mathbb C$-case, for a $\mathbb Q$-function $F(Z)$ to depend only on the $\mathbb Q$-variable $Z$ (but not on its conjugate $Z^*$, etc.), one arrives at a strongly overdetermined PDE-system, which is, in fact, incompatible~\cite{Sudbery}.
 
On the other hand, the derivative of a $\mathbb Q$-function is, in fact, indefinite, being dependent on the direction of an infinitesimal increment of $\mathbb Q$-variable, contrary to the most important property of $\mathbb C$-holomorphic functions. 
As for physical applications, all the currently known approaches do not lead to some principally novel predictions. 

Meanwhile, an alternative approach to $\mathbb Q$-analysis had been proposed in our works (see, e.g.,~\cite{AD3} and references therein). 
This is based on the old construction of G. Sheffers~\cite{Sheff} on the analysis over a {\it commutative associative algebra} $\mathbb A$, which in the $\mathbb C$-case reduces to the canonical theory, yielding the Cauchy--Riemann holomorphic conditions in particular. 

\textls[-15]{Following Sheffers, an $\mathbb A$-differentiable function $F(Z),~Z\in \mathbb A$ should satisfy the condition}
\be{sheff}
 dF = G \cdot dZ \;, 
\ee
where $G=G(Z) \in \mathbb A$ is an auxiliary $\mathbb A$-function, which, in the case of the division algebra $\mathbb C$, can be identified with the derivative $G=F^\p (Z)$. 
Writing out \eqref{sheff} in components, one obtains 
\be{diff1}
 \prt_a F^d = C^d_{ab} G^b \;,
\ee
where $\{C^d_{ab}\},~a,b,\dots=1,2,\dots, N=dim \mathbb A$ are the structure constants of $\mathbb A$-algebra. 
If one demands, as in $\mathbb C$, the ``derivative'' $G(Z)$ be $\mathbb A$-differentiable as a consequence of \eqref{diff1}, 
\be{diff2}
 \prt_a G^d = C^d_{ab} H^b \;,
\ee
\textls[-15]{then the commutator $\prt_{[a b]} F^d =0$ vanishes identically solely due to the constraints of associativity and commutativity on the structure constants. 
Thus, the systems \eqref{diff1} and \eqref{diff2} are self-compatible, and the corresponding function $F(Z)$, if differentiable once, is infinitely differentiable, hence $\mathbb A$-analytical. 
Evidently, in the $\mathbb C$-case, after eliminating the $G$-components from \eqref{diff1} the resulting conditions on the derivatives of $F$ coincide with the Cauchy--Riemann equations.  }

It is worth noting that the approach suggested by Sheffers has been successfully applied to construct the so-called ``superanalysis'', that is, the analysis over the direct sum of 4D commutative and Grassmanian space--time algebras~\cite{Volovich1,Volovich2}. 
 
Now, definition \eqref{sheff} can be naturally generalized to the case of a {\it noncommutative} associative algebra, Hamilton's quaternions $\mathbb Q$ in particular. 
Specifically, one requires a $\mathbb Q$-valued function $F(Z)\in \mathbb Q$ to be {\it $\mathbb Q$-differentiable}~\cite{AD3}, 
\be{kass}
 dF = \Phi \cdot dZ \cdot \Psi \;, 
\ee
where $\Phi=\Phi(Z),~\Psi=\Psi(Z)$ are now two auxiliary $\mathbb Q$-valued functions that can be named ``semi-derivatives'' (left and right, respectively). 
Note that these are defined up to an element $\alpha(Z)$ of the center of $\mathbb Q$ (that is, a real-valued function of $\mathbb Q$-variable). 
 
Thus, {\it a $\mathbb Q$-valued function is $\mathbb Q$-differentiable if its differential $dF$ can be represented in a component-free form, that is, via the operation of multiplication in $\mathbb Q$}. 
Evidently, in the case of a commutative associative algebra definition, \eqref{kass} reduces to Sheffer's condition \eqref{sheff}, $dF \mapsto (\Phi \cdot \Psi) \cdot dZ \equiv G \cdot dZ$. 
 
As in the case of a $\mathbb C$-holomorphic function, any mapping in the 4D Euclidean space $\bf E^4 \mapsto E^4$ realized by a $\mathbb Q$-differentiable function $F(Z)$ is a {\it conformal} one, since one has \mbox{from \eqref{kass}{:}} 
\be{conform}
 N(dF) = N(\Phi) N(dZ) N(\Psi) = N(\Phi \cdot \Psi) N(dZ) \equiv \Lambda(Z)N(dZ) \;, 
\ee
where $N(Z)\in \mathbb R$ is the positive definite norm of a quaternion $Z$, and $\Lambda(Z)\in \mathbb R$ is the corresponding conformal scale factor. 
              
The explicit connection between $\mathbb A$-differentiable functions and conformal mappings reveals one more analogy between complex analysis and noncommutative $\mathbb Q$-analysis defined by \eqref{kass}. 
However, it is well known that the group of conformal mappings in $\bf E^4$ is 15-parametric, contrary to the infinitely dimensional conformal group of the complex plane. 
Conversely, it is easy to prove~\cite{AD3} that any of such conformal mappings can be realized by a $\mathbb Q$-differentiable function $F(Z)$ defined by \eqref{kass}. 
For example, an {\it inversion} w.r.t. a sphere $S^3 \in \bf E^4$ is algebraically  
represented by the function $F(Z)=Z^{-1}$ whose differential 
\be{invers}
 dF = dZ^{-1} = - Z^{-1} dZ Z^{-1} \;,
\ee               
has the form corresponding to \eqref{kass}. 

On the other hand, the class of $\mathbb Q$-differentiable mappings \eqref{kass} is therefore too narrow to be used for physical applications. 
Fortunately, the situation drastically changes when one passes to the complex generalization of $\mathbb Q$, the algebra of biquaternions $\mathbb B$. 

Indeed, in the $\mathbb B$-case one can consider the semi-derivatives $\Psi$ (or $\Phi$) in \eqref{kass} whose (complex-valued) norm is null, say, $N(\Psi)=0$, so that the function $\Psi(Z)$ is a {\it null divisor}. 
Evidently, in this case the corresponding mapping $Z \mapsto F(Z)$ on $\bf C^4$, vector space of $\mathbb B$, {\it is not a conformal mapping}. 
Therefore, one expects that the class of $\mathbb B$-differentiable mappings is rather wide and 
the corresponding functions $F(Z),~Z\in \mathbb B$ subject to \eqref{kass} can be treated as fundamental physical fields. 

Biquaternions ($\mathbb B$) do not form an exceptional algebra just because of the existence of elements-null divisors. 
In fact, the algebra $\mathbb B$ is isomorphic to the full $2\times 2$ matrix algebra over $\mathbb C$,~~$\mathbb B$$\sim$$Mat(2,\mathbb C)$. 
Moreover, its automorphism group $SO(3,\mathbb C)$ is isomorphic, $SO(3,\mathbb C)$$\sim$$SL(2,\mathbb C)$, to the six-parametric spinor Lorentz group $SL(2,\mathbb C)$. 
That is why the $\mathbb B$-algebra looks like a natural candidate to be {\it the space--time algebra}! 

On the other hand, the vector space of $\mathbb B$ is eight-dimensional (over the reals); so, one has to deal with the notorious interpretation problems of the four extra coordinates. 
Moreover, the Minkowski space $\bf M$ does not even form a subalgebra of $\mathbb B$. 
However, the first version of {\it algebrodynamics} has been successfully developed under the assumption that the coordinate space in \eqref{kass} is restricted to the Hermitian matrices $Z \mapsto X=X^+$, that is to the Minkowski space $\bf M$. 
In the procedure, however, the fields $F(X),\Phi(X),\Psi(X)$ are still complex-valued, as is often the case in physics. 
 
Besides, below, we restrict ourselves to the simplest and most remarkable case when one of the ``semi-derivatives'', say, $\Psi=\Psi(X)$ coincides with the principal function $F(X)$. 
Thus, in the framework of algebrodynamics, we consider the following reduced form of $\mathbb B$-differentiability conditions \eqref{kass}:
\be{GSEF}
 dF = \Phi \cdot dX \cdot F \;,
\ee
\textls[-15]{where $X=X^+ = x^0 {\bf 1} + x^{a} {\bf \sigma}_{a}$,~~$\{x^\mu\},~\mu=0,1,2,3$ are the Minkowski space--time coordinates, while $\bf 1$ and $\{{\bf \sigma}_{a}\},~a=1,2,3$-the $2\times 2$ unit matrix and three Pauli matrices, respectively. }
  
Since, in the full $2\times 2$ matrix representation of $\mathbb B$, both columns of $F(X)$ are completely independent, one can compose any solution of \eqref{GSEF} of solutions to the following \linebreak reduced equation:
\be{GSES}
 d\xi = \Phi dX \xi \;, 
\ee
where $\xi=\xi(X)$ is one of the columns of the matrix $F(X)$. 
Hereinafter, system \eqref{GSES} is referred to as {\it the generating system of equations} (GSE). 

Note that the GSE \eqref{GSES} is Lorentz invariant. 
This is because, under global transformation of coordinates from the proper Lorentz group, 
\be{Lortrans}
 X \mapsto S^+ X S,~~S \in SL(2,\mathbb C) \;, 
\ee   
the column $\xi(X)$ behaves as a two-spinor, 
\be{spinor}
 \xi \mapsto S^{-1} \xi
\ee
while the matrix $\Phi$ transforms as a (complex) four-vector
\vspace{6pt}
\be{4vec}
 \Phi \mapsto S^{-1} \Phi (S^{+})^{-1} \;. 
\ee
           
The reduced conditions of $\mathbb B$-differentiability \eqref{GSEF} or, equivalently, \eqref{GSES} constitute a full set of equations of an {\it algebraic field theory}, the so-called ``algebrodynamics''~\cite{AD3}. 
Such a theory is Lorentz invariant and contains a natural gauge and spinor (twistor) structure (see below). 
It contains a set of fundamental relativistic fields whose free equations do hold identically on the solutions of \eqref{GSES}. 
Note that the latter is over-determined, and both the spinor $\xi(X)$ and the vector $\Phi(X)$ fields can be found from it (see below). 

For this reason, the system of PDEs corresponding to \eqref{GSES} is non-Lagrangian. 
Nonetheless, a complete set of conservation laws holds on the solutions of \eqref{GSES} so that the related algebraic dynamics is, in fact, {\it conservative}. 
As for particles, {\it one can identify them with (isolated) singularities} of the primordial fields $\xi(X),\Phi(X)$ or secondary gauge fields, which can be defined via the latter. 
Other remarkable properties of the principal GSE \eqref{GSES} and its solutions will be considered below. 
  
\section{Twistor Structure and General Solution to the Equations of Biquaternionic Differentiability}\label{sec3}

We have seen that, in the simplest and most remarkable case, equations of $\B$-differentiabi-~\linebreak~lity reduce to the following (GSE) form \eqref{GSES}:
\be{GSEN}
 d\xi = \Phi dX \xi \;.
\ee

Rewriting system \eqref{GSEN}, one obtains:
\be{link}
 d\xi=\Phi d(X\xi) - \Phi X d\xi, \rightarrow (I+\Phi X) d\xi = \Phi d\tau \;, 
\ee
where the two-spinor 
\be{incid2}
 \tau:= X\xi
\ee
together with the primary two-spinor $\xi$ forms a {\it twistor} on the Minkowski space--time $\bf M$ (we ignore in the incidence relation \eqref{incid2} the generally accepted factor $i$, cp. with \eqref{inc}). 

Relation \eqref{link} indicates that the four twistor components, that is, the two two-spinors $\{\xi,\tau\}$, are {\it functionally dependent}. 
Precisely, for any solution of \eqref{GSEN}, there exist two independent implicit functions $\{\Pi^C\},~ C=1,2$ of four complex arguments such that one has 
\be{Gensol}
 \Pi^C (\xi,\tau) = 0 \;.
\ee

In fact, the two algebraic equations \eqref{Gensol} represent the {\it general solution} to the equations of the reduced (GSE) form of $\B$-differentiability conditions. 
Indeed, at any point $X \in \bf M$, the two equations \eqref{Gensol}, that is, 
\be{GensolV}
 \Pi^C (\xi, X\xi)= 0 \;,
\ee
can be resolved w.r.t. the two unknowns $\{\xi^A\},~A=0,1$. 
Executing the procedure at every point $X$ and selecting a continuous branch of the roots of \eqref{GensolV}, one obtains a spinor field $\xi(X)$. 
Generally, such a {\it multi-valued} field admits a whole set of continuous branches. 

Substituting one of the branches of the solution into \eqref{GensolV}, one has an identity that can be differentiated w.r.t. any of the coordinates $X$. 
Below, however, we shall return back to consider the complexification $X \rightarrow Z=\{Z^A_B\}$, which corresponds to the whole vector space $\C\bf M$ of the $\B$-algebra. 
The identity \eqref{GensolV} should be rewritten in the form
\be{GensolZ}
 \Pi^C (\xi^A, Z^A_B \xi^B)=0 \;,
\ee
which, after differentiation w.r.t. a complex coordinate $Z^B_D$, gives
\be{diff}
 0=\frac{d\Pi^C}{dZ^B_D}=P^C_A (\prt^D_B \xi^A)+\frac {\prt\Pi^C}{\prt \tau^B} \xi^D \;,
\ee
where 
\be{total}
 P^C_A:=\frac{d\Pi^C}{d\xi^A} =\frac {\prt \Pi^C}{\prt \xi^A} + \frac{\prt \Pi^C}{\prt \tau^E} Z^E_A \;, 
\ee
and the evident notation is used: $\prt^D_B: = \prt/\prt Z^B_D$. 
   
At the points where the matrix $Q^E_C$ inverse to $P^C_A$ exists, 
\[ 
Q^E_C P^C_A = \delta^E_A \;,
\]
one finds from \eqref{diff}
\be{NSF}
 \prt ^D_B \xi^E = -\frac{d\Pi^C}{d\tau^B} Q^E_C \xi^D \;, 
\ee
and multiplying \eqref{NSF} by $\xi_D$ obtains the following simple relation
\be{SFCA}
 \xi_D \prt ^D_B \xi_E =0 \;. 
\ee

The latter, after multiplication by $\xi^E$, represents the (complexified over $\C\bf M$) well-known defining equations~\cite{Penrose1}, ch. 6) of a null shear-free congruence (NSFC),
\be{SFC}
 \xi^E \xi_D \prt ^D_B \xi_E =0 \;.
\ee

Remarkably, the latter are {\it projectively invariant} so that only the {\it ratio} of two spinor components $\xi^A$ can be found from them. 
On the contrary, the original equations \eqref{SFCA} is more rigid and define {\it both components} of the spinor field! 
 
The principal equations \eqref{SFCA} can also be obtained (after multiplication by $\xi_D$) from the original GSE \eqref{GSEN}, which, in index notation, has the form
\be{GSEI}
 \prt_B^D \xi^E = \Phi_B^E \xi^D \;.
\ee

It is now evident that the ``orthogonality'' of the derivatives of any two twistor components follows immediately, 
\be{ortho}
 \prt^A_D \xi_B \prt^D_A \xi_C = \prt^A_D \xi_B \prt^D_A \tau_C = \prt^A_D \tau_B \prt^D_A \tau_C =0 \;.
\ee

On the other hand, on the solutions of \eqref{GSEN}, any two twistor components are always functionally independent, while the third one depends on them due to the (projectively invariant) structure of general solution \eqref{GensolV}, see, e.g.,~\cite{IJGMMP}.     

To summarize, we see that {\it any $\B$-differentiable function (from the principal GSE class (\ref{GSEN})) defines an NSFC on the Minkowski space--time $\bf M$ as well as on its complexification $\C \bf M$}.
        
Consider now the points at which the determinant 
\be{sing}
 \det \vert\vert P^C_A\vert\vert \equiv \det \vert\vert \frac{d\Pi^C}{d\xi^A} \vert\vert=0
\ee
vanishes. 
These correspond to the {\it merging} of a pair of branches of the multi-valued field $\xi(Z)$ or, equivalently, to {\it multiple} roots of the generating algebraic system \eqref{GensolZ}. 
At such singular points, the derivatives $\prt^D_B \xi^A$, in view of \eqref{total}, become infinite. 
We shall see below that the gauge fields associated with the solutions to \eqref{SFCA} or \eqref{SFC} are also singular there. 

Consider further the twistor space $\{\xi,\tau\}$ itself and assume that, in the corresponding domain, a continuous branch of the function $\tau= \tau(\xi)$ can be determined from two implicit equations  \eqref{Gensol}. 
At these points, the matrix of derivatives $\prt\Pi^C /\prt \tau^B$ is nonsingular and together with the definition \eqref{total}, allows us to transform the condition of singular locus \eqref{sing} into the following form:  
\be{locus}
 \det \vert \vert \wh Z^B_A - Z^B_A \vert\vert =0 \;,
\ee
where 
\be{genrtstring}
 \wh Z^A_B:= - \frac{\prt \Pi^C}{\prt \xi^A} R^B_C
\ee
and $R^B_C$ is the inverse matrix, $(\prt \Pi^C/\prt \tau^E) R^B_C = \delta^B_E$. 

Generally, points of the singular locus on $\C\bf M$ are determined by one complex condition \eqref{locus} and, therefore, compose a {\it complex hypersurface}. 
However, on the Minkowski cut $\bf M$, there are two constraints on the four real space--time coordinates $X=\{x_\mu\}$ represented by real and imaginary parts of \eqref{locus}. 
Therefore, the {\it singular locus of an NSFC on $\bf M$ is generically a string} whose worldsheet can be parametrized by two real parameters, say, 
\be{string}
 X = X(\rho,\sigma) \;.
\ee

It also follows from \eqref{locus} that any singular point $X\in \bf M$ belongs to a null complex cone of a point on the {\it generating complex string} $\wh Z(\rho,\sigma)$, which can be also parameterized by two real parameters $(\rho,\sigma)$ and, in particular, ``moves'' through ${\mathbb C}\bf M$ in ``real time'' $\rho$. 
 
\section{Gauge Symmetry and Fields Associated with Solutions of $\mathbb B$-Differentiability Conditions}\label{sec4} 

The physical fields defined by the $\mathbb B$-differentiability conditions, and therefore by NSFC, have been described in detail in a number of works, see, e.g.,~\cite{IJGMMP}. 
Below, we briefly review this issue and supplement it with a second type of Maxwell-like field, which can be associated with any NSFC. 

Consider again the reduced form GSE of $\mathbb B$-differentiability conditions \eqref{GSEN}. 
\mbox{System \eqref{GSEN}} is overdetermined and its {\it conditions of compatibility} can be easily obtained~\cite{IJGMMP}) 
\be{compat}
 dd\xi \equiv 0 = R\xi,~~R:=d\Phi - \Phi \wedge dX \wedge \Phi \;, 
\ee    
where $\wedge$ denotes the wedge product. 
From \eqref{compat}, it {\it does not} follow that the effective $2\times 2$ matrix curvature $R$ is zero. 
Instead~\cite{selfdual}, it is {\it (anti)self-dual}. In particular, if one identifies the matrix $\Phi(X)=A^\mu \sigma_\mu,~~\sigma_\mu:=\{\bf 1, \sigma_a\}$ with the four-potentials $A^\mu(X)$ of a (complex) electromagnetic field, then the corresponding field strength tensor 
\be{emfield}
 F_{\mu\nu}:=\prt_\mu A_\nu - \prt_\nu A_\mu
\ee
satisfies the equations 
\be{selfdual}
 F_{\mu\nu} = \frac{i}{2} \varepsilon_{\mu\nu\rho\lambda} F^{\rho\lambda} \;.
\ee

In the usual manner, free (complexified) Maxwell equations follow from the (anti)self-duality constraint \eqref{selfdual}. 
Moreover, it can be proved~\cite{Sing} that any isolated singularity of the corresponding Maxwell field is either electrically neutral or possesses an effective electric charge~(defined as the flux of electric field through a closed surface encircling the singularity), which is a {\it multiple integer} of a minimal one associated with the static axisymmetric Kerr-like congruence (see below, Section \ref{sec5}). 
Thus, the electric charge of an electromagnetic field defined from the ``master equations'' \eqref{GSEN} {\it is necessarily self-quantized}~\cite{Sing}. 

The electromagnetic interpretation of the components $\Phi(X)$ is also supported by a gauge-like symmetry of the GSE equations \eqref{GSEN} and their general solution \eqref{Gensol}. 
Specifically, the latter, together with the incidence relation \eqref{incid2}, are invariant under the following transformation of the involved twistor field:
\be{gauge}
 \xi \mapsto \alpha(\xi, \tau) \xi,~~~ \tau \mapsto \alpha(\xi, \tau) \tau \;,
\ee
where the function $\alpha \in \mathbb C$ should depend on the coordinates $X$ {\it only implicitly}, through the initial twistor field $\xi(X),\tau(X)$ under transform. 
It is easy to demonstrate that, together with \eqref{gauge}, the components of potentials $\Phi(X)$ should transform in a gradient-wise way, 
\vspace{6pt}
\be{grad}
 \Phi_{AA^\p} \mapsto \Phi_{AA^\p} - \prt_{AA^\p} \alpha \;, 
\ee
\textls[-15]{in order that the set of GSE equations be form invariant. 
The symmetry represented by the combined transformations \eqref{gauge} and \eqref{grad} has been called ``weak'' or restricted gauge invariance. 
These transformations constitute a {\it proper subgroup} of the full gauge group $\mathbb C$~\cite{IJGMMP}.  } 

In fact, the electromagnetic field is defined by the {\it trace} part of the effective curvature $R$, while its {\it trace-free} part can be associated with a {\it complex $SL(2,\mathbb C)$ Yang--Mills field}. 
In view of \eqref{selfdual}, the corresponding curvature satisfies the free Yang--Mills equations~\cite{IJGMMP}. 
Finally, from the four components of matrix $\Phi(X)$, one can compose two pairs, any of which {\it satisfies the two-spinor Weyl equations}~\cite{IJGMMP}.
 
As for the principal spinor field $\xi(X)$, we have seen that both its components 
are subject to the {\it complex eikonal equation}. 
In addition, their {\it ratio} (which defines the structure of corresponding NSFC) on any solution to \eqref{GSEN} satisfies the {\it linear wave equation}~\cite{Wilson, Sing}.

We now demonstrate that there exists a second structure, which can also be identified with an electromagnetic field. 
Consider two functionally independent components $\alpha(X), \beta(X)$ of the principal twistor field $\{\xi,\tau\}$. 
Apart from the complex eikonal equation 
\be{eiktwo}
 \prt_\mu \alpha \prt^\mu \alpha =0,~~\prt_\mu \beta \prt^\mu \beta =0 \;,
\ee
the components satisfy, as a consequence of the orthogonality conditions \eqref{ortho}, an \linebreak additional constraint
\be{mix}
 \prt_\mu \alpha \prt^\mu \beta =0 \;.
\ee
   
Let us now define the field strength differential two-form $C$ as follows:
\be{secstr}
 C=d \alpha \wedge d \beta \;. 
\ee

It is clear that $C=d(\alpha d \beta)$, so that 
\be{dualstr}
 dC=0
\ee
and the first pair of Maxwell equations is satisfied trivially. 
{Note, however, that since $\alpha(X)$ and $\beta(X)$ are functionally independent components of the twistor $\{\xi,\tau\}$, the one-form $\alpha d \beta$ is not a pure gauge and the field strength $C$ is non-trivial (for a concrete example, see \mbox{Section \ref{sec5})}}.

On the other hand, it is easy to see that, in view of \eqref{eiktwo} and \eqref{mix}, one has 
\be{batecond}
 \star(d\alpha \wedge \star C) = \star(d\beta \wedge \star C) = 0,
\ee 
where $\star$ denotes the Hodge dual. 
This gives $d\alpha \wedge \star C=0$ and $d\beta \wedge \star C=0$, which together imply that $\star C$ is a multiple of $d \alpha \wedge d \beta$, hence, proportional to $C$. 
Since $\star^2 = -1$, we deduce (compare with~\cite{Bateman}, p. 12) 
\be{selfdual2}
 \star C = \pm i C \;,
\ee
{i.e., the field strengths are (anti-)self-dual (for recent work on self-dual fields see, e.g.,~\cite{symdual,symdual1}),}. 
Therefore, the second pair of Maxwell free equations $d \star C = 0$ follows as an immediate consequence of the first \eqref{dualstr}. 
It is also easy to see that 
$C_{\mu\nu}C^{\mu\nu} =C_{\mu\nu}\tilde C^{\mu\nu}= \tilde C_{\mu\nu}\tilde C^{\mu\nu}=0$, so that the introduced Maxwell-like field is null (here, $ \tilde C_{\mu\nu}$ denote the components of $\star C$). 
   
To conclude, there are a lot of fundamental relativistic (massless) fields that  can be defined through the quantities present in the structure of the reduced form \eqref{GSEN} of $\mathbb B$-differentiability conditions or, equivalently, through the structure of NSFC. 
Moreover, the corresponding free field equations identically hold for the solutions of \eqref{GSEN}. 
However, not all these fields have direct physical meaning, but just those that govern the behavior of particle-like formations (isolated caustics-singularities). 
This important problem requires further study. 

\section{Solutions, Associated Fields and Singularities}\label{sec5}

Let us briefly review now the simplest known solutions to the reduced form \eqref{GSEN} of $\mathbb B$-differentiability equations \eqref{kass} together with the fundamental fields, which can be defined via the latter, along with an interesting novel example of the second type \eqref{secstr} of a \mbox{Maxwell-like field.} 

All such solutions can be obtained in a purely algebraic way by the corresponding selection of the generating twistor functions in \eqref{Gensol}. 
Since all the related fields depend, in fact, only on the ratio $G=G(X)$ of two spinor components, we can reduce the general solution \eqref{GensolV} to a single algebraic equation of the form
\be{proj} 
 \Pi(G, \tau^1,\tau^2)=0,~~\tau_1:=wG+u,~~\tau_2:=vG+\bar w \;,
\ee
where $\tau_1, \tau_2$ are two projective twistor components while $u,v=t\pm z,~\bar w$, \linebreak $w =x\pm iy$-the spinor coordinates on $\bf M$. 
In particular, the only static spherically symmetric (w.r.t the corresponding NSFC and associated fields) solution is obtained from the following generating function:
\be{stat}
 \Pi=G \tau_1 - \tau_2 = wG^2 +2zG -\bar w \;.
\ee
Resolving the equation $\Pi=0$ for the ratio $G$, one obtains a two-valued distribution
\be{stereo}
 G =\frac{\bar w}{z\pm r},~~r:=\sqrt{x^2+y^2+z^2},
\ee
which geometrically represents the {\it stereographic projection} $S^2 \mapsto \mathbb C$. 

Making use of \eqref{emfield} and calculating now the effective four-potentials and strengths of the associated electromagnetic field, one obtains the Coulomb-like distribution~\cite{AD3} with the following electric $\bf E$ and dual magnetic $\bf H$ fields:
\be{coulomb}
 E_r:=\pm \frac{q}{r},~~H_r=\pm\frac{iq}{r},~~ q=1/4 \;,
\ee
the other components being null. 
The electric charge is fixed, being equal in modulus to the imaginary magnetic charge and can be identified (after passing to dimensional units) with the {\it elementary charge}. 

The associated solution to Weyl equations, which for two components $\psi_1,\psi_2$ of the two-spinor can be represented in the following form:
\be{weylform}
 \prt_{\bar w} \psi_1 = \prt_u \psi_2,~~\prt_v \psi_1 = \prt_w \psi_2 \;,
\ee 
is given by the following ansatz~\cite{Sing}:
\be{weyl}         
 \psi_1 = -\frac{\bar w}{r(z+r)},~~\psi_2 = \frac{1}{r} \;,
\ee
the two-spinor components being composed from corresponding components of the four-potential matrix $\Phi(X)$. 

Let us now compute the strengths of the ``second'' electromagnetic field \eqref{secstr}. 
For this, we choose the following two functionally independent components of the projective twistor entering \eqref{stat}:
\be{twistcomp}
 \alpha = G = \frac{\bar w}{z+r},~\beta:=\tau_1:= t+r 
\ee
and obtain the following expression for the field strengths ${\bf C}:={\bf E} + i{\bf H}$ in \linebreak spherical coordinates:
\be{screw}
 {\bf C}_\theta = \frac{1}{r \cos^2 \theta/2} e^{i\varphi},~~{\bf C}_\varphi =i {\bf C}_\theta,~~~{\bf C}_r =0 \;.
\ee

\begin{figure}[h]
\includegraphics[width=12cm, trim = {4cm 6cm 6cm 4cm}, clip]{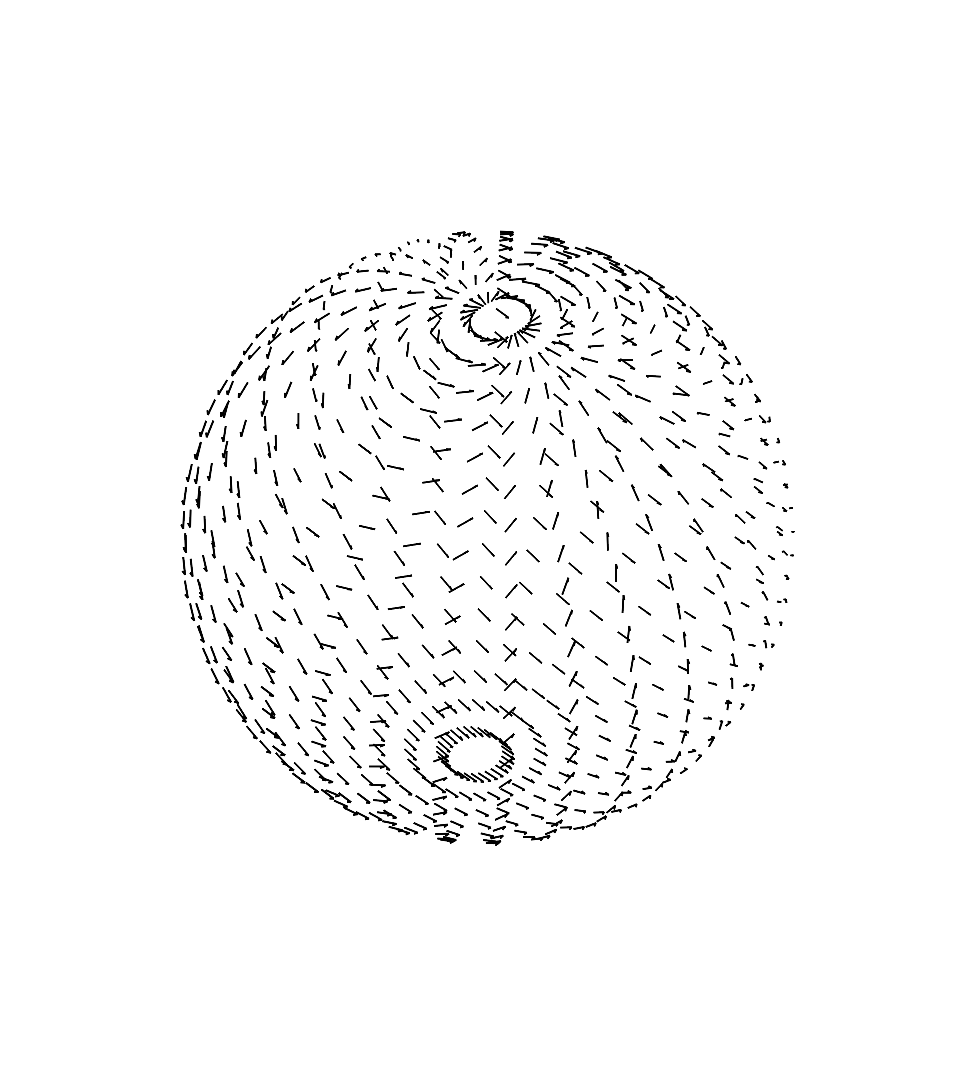}
\caption{The electric field polarization pattern on a sphere.}
\label{nulrot}
\end{figure}

One observes that the free electromagnetic field is static despite the fact that both invariants ${\bf E}^2 - {\bf H}^2$ and ${\bf E}\cdot {\bf H}$ are identically null. 
The singular locus consists of the central point $r=0$ and the half-axis $-\infty<z<0$. 
For any $r$, the field is tangent to the corresponding two-sphere (see Figure~\ref{nulrot}) and resembles the action of the so-called null rotation (see, e.g.,~\cite{Penrose2}, p. 29). 
Moreover, this static solution ${\bf E}$, {\bf H} can be promoted to a spherical wave: ${\bf E}f(r-t)$, ${\bf H}f(r-t)$ and $-{\bf E}f(r+t)$, ${\bf H}f(r+t)$, where $f$ is an arbitrary smooth function of its argument.
\vspace{1pt}

A number of more complicated solutions to the $\mathbb B$-differentiability equations \eqref{GSEN}, with their corresponding NSFC and loci of singularities-caustic,s can be found in~\cite{Sing}. 

It is also well known that NSFC give rise to the so-called Kerr--Schild metrics 
\be{Shild}
 g_{\mu\nu} = \eta_{\mu\nu} + H(X) k_\mu k_\nu \;, 
\ee
$H(X)$ being a scalar field (``gravitational potential''). 
For a large class of the NSFC, one manages to fit the latter so that $g_{\mu\nu}$ is a solution of the vacuum Einstein equations (or, together with the appropriate four-potential field,-to electrovacuum Einstein--Maxwell system). 
In such a way, the metric corresponding to principal solution \eqref{stereo} is certainly the Schwarzchild one. 
It is noteworthy that, under a complex shift $z \mapsto z+ia,~~a \in \mathbb R$, the congruence transforms into the Kerr one, with singular locus of the form of a ring of radius $a$. 
The associated electromagnetic field and Riemannian metric become identical to those of the Kerr--Newman solution.     

\section{Conservative Algebraic Dynamics on the Unique Worldline}\label{sec6}

\textls[-15]{Below, we shall briefly review the content of papers~\cite{Ildus1, Ildus3} on a peculiar {\it algebraic} realization of the old idea of J. A. Wheeler and R. P. Feynmann~\cite{Feynman1, Feynman2} related to unified description of interacting particles. 
They assumed that all the primary particles in the Universe are in fact one and the same point particle located at different points on the Unique Worldline (UWL). 
That is why, this construction received the informal name, ``One-Electron Universe''. Earlier~\cite{Stueckel1,Stueckel2}, similar ideas have been advocated by E. C. G. Stueckelberg.}

The UWL concept opens a lot of attractive advantages concerning, in particular, the possibility to naturally explain the property of identity of ``electrons'', etc., and to fix the true form of particles' interaction. 
However, a number of fundamental drawbacks, e.g.,~difficulties with the problem of conservation laws, did not allow the UWL concept to receive essential development and recognition. 

Meanwhile, in~\cite{Ildus1}, a non-relativistic version of the Wheeler--Feynman concept has been elaborated. 
We assumed that the UWL can be defined {\it implicitly} by a system of three algebraic equations
\be{nrel}
 F_a(t,x,y,z)=0 \;, 
\ee
where $\{F_a\},~a=1,2,3$ are three independent functions of three Cartesian coordinates $x_a =\{x,y,z\}$ and the time parameter $t$. 
For a number of reasons~\cite{Ildus1}, we confine ourselves to {\it polynomial} functions $\{F_a\}$.

At any moment $t$, {\it the roots of algebraic system \eqref{nrel} correspond to a spacial distribution of a set of identical point particle-like formations, which, in the course of time $t$, ``move'' in concord along the UWL}. 
Thus, one encounters here a peculiar form of collective algebraic (non-Lagrangian) dynamics, realizing the Wheeler--Feynman idea in a simple way.

There exist, in fact, two kinds of identical particle-like formations, represented by real (R) or complex conjugate (C) roots of the system \eqref{nrel}. 
At a number of instants $t=\{t_k\},~ k=1,2, ...$, a pair of real roots \eqref{nrel} may merge, transforming further into a pair of complex conjugate roots. 
This transformation can be naturally identified with the {\it process of annihilation} of two R-particles along with their transformation into a pair of C-particles. 
Certainly, the inverse {\it process of pair production} at some other instants can also occur. 
{Moreover}, the {\it trajectory} of motion defined by \eqref{nrel}, after elimination of the time parameter, consists generally of a {\it number of isolated curves} in the 3D space together with those connecting them in complexified space. 
Typically, two R-particles can annihilate. The emerging C-particles propagate in C-space for some time before they give rise to a new pair of R-particles on another branch of the UWL~\cite{Ildus1}. 

An amazing property of the collective algebraic dynamics is its conservative character. Precisely, for any non-degenerate polynomially parametrized UWL, {\it the laws of conservation of total momentum, angular momentum and (the analogue of) total energy hold. Remarkably, this is a direct consequence of the Vieta's formulas for the roots of polynomial system \eqref{nrel}} (for details, see~\cite{Ildus1}).

\section{Relativistic Algebraic Dynamics on the Unique Worldline}\label{sec7}
       
It is relatively straightforward to pass to a relativistic generalization of the UWL concept~\cite{Ildus3} based on the concept of the {\it local light cone} of a particle and, in its simplest realization, this does not even require the UWL be defined {\it implicitly}. 
Remarkably, such a {\it relativistic generalization is closely related to the algebrodynamical concept, and the equations of $\mathbb B$-differentiablility in particular}.

Specifically, let us consider a (considerably complicated) worldline, which can be parameterized in the usual manner, as $x=\{x^\mu (\sigma)\}$, with a monotonically increasing timelike parameter $\sigma$. 
It is well known~\cite{Cramer} that the congruence of straight ``light-like'' rays radiated by the particle is necessarily null shear-free (NSFC).

Indeed, let a matrix $X=X^+=\{X^\mu \}$ represent a point of observation. The expression for the null interval, that is, the {\it light cone equation} (LCE)
\be{null}
 \det \Vert X - x(\sigma) \Vert =0 \;, 
\ee
allows one to define the projective twistor field $\{ \xi,~ \Pi:=X\xi \}$, 
\be{twfield}
 \Pi -\pi(\sigma):= (X - x(\sigma))\xi =0 \;. 
\ee

This is a pair of linear (w.r.t. $\xi$) equations, which shows that the three components of the projective twistor, $G:=\xi^2/\xi^1, \Pi^1, \Pi^ 2$, are functionally dependent. 
Therefore, the congruence under consideration is necessarily an NSFC, and the corresponding spinor-twistor field is, in fact, a solution to the reduced equations of $\mathbb B$-differentiability \eqref{GSEN}.

Consider now an inertially moving point-like (idealized) observer O, $X=X(\tau)$. The points on the UWL that can be instantaneously ``detected'' by O are defined by the LCE \eqref{null}.  

At any $\tau$, one has, generally, a lot of (real R or complex conjugate C) roots $\{ \sigma_k \}$ that define a whole ensemble of point-like (R- or C-) particles on the UWL or its complex extension. 
Such identical particle-like formations have been called {\it duplicons }~\cite{Dupl}. 
 
For the observer, it is not important whether the UWL is completely real or complex-valued: in any case, he/she {\it instantaneously} receives a number of light-like signals from different points on the same UWL. 
On the other hand, at a discrete set of time instants, a pair of duplicons merges, as the corresponding roots of \eqref{twfield} become multiple. 
At these moments, one has an amplification (caustic) of the twistor field along a null straight ray connecting the points of merging and observation. 
Such a null structure propagating in three-space with the speed of light can be regarded as a {\it classical model of the photon}.             

From this viewpoint, the location of a ``particle'' can be detected by O only at particular instants of time, while the duplicons themselves are not permanently observable. 
That is why one can speak about ``dimerous electron''~\cite{Dupl}, which, in fact, {\it exists only instantaneously}, while permanently it ``exists'' as divided into two identical halves---duplicons. 
This concept can elegantly explain the phenomenon of {\it quantum interference}~\cite{Dupl}, replacing (probability) waves' superposition of quantum theory with the shift of phase of ``complex time''. 

Let us finally enumerate some of the most remarkable consequences of the relativistic UWL construction. 
First of all, for polynomially parametrized non-degenerate UWL, one again observes {\it a complete set of conservation laws are fulfilled in a Lorentz invariant form}. 
For example, the law of conservation of total angular momentum can be written as 
\be{angmom}
 \Sigma ({\dot x}^\mu x^\nu - {\dot x}^\nu x^\mu) = const \;, 
\ee
where the summation ranges over the coordinates and velocities of all of the duplicons-particles of equal mass that can be set to unity. 
However, {\it the dynamics is conservative only for uniformly moving observers}~\cite{Ildus3}, this property being in full agreement with the canonical principles of physics.
 
For a timelike polynomially parametrized UWL, there are also some rather unexpected properties of the behavior of duplicons at large values of the proper time of the observer. 
Specifically, at some critical time $T_1$, the effect of {\it coupling} of duplicons (which asymptotically approach each other) takes place. 
At a greater time $T_2 > T_1$, the almost-formed pairs start to approach one another so that a set of scattering {\it clusters} containing a great number of duplicons' pairs emerges. 
The two effects, formation of pairs and clusterization, on the background of the expanding ``Universe of roots-particles'' strongly resembles, at least qualitatively, the real physical pattern. 
 
\section{Conclusions}\label{sec8}

The paper presents an overview of the tenets and main consequences of algebrodynamics (AD). 
To a certain extent, this theory is unique, since it is essentially based on a single relation \eqref{kass}, a generalization of the Cauchy--Riemann holomorphy conditions to the algebra of complex quaternions (biquaternions) $\mathbb B$, the direct sum of two exceptional algebras. 
Algebraic structures of other dimensions and signature currently attract considerable attention as possible candidates for space--time algebra, see, e.g.,~\cite{Duff2019, Nishino1996, Bars1998, popmat1}. 
Interpreting $\mathbb B$-differentiable functions as physical fields, and particles as their singularities, relation \eqref{kass} (more precisely, its reduced form \eqref{GSEN}) implies a complete and self-consistent theory of interacting fields/particles. 
{Remarkably, the over-determined} nature of the system of equations of $\mathbb B$-differentiability leads to the “selfquantization” of the associated \linebreak electric charge. 

Furthermore, many structures accepted in theoretical physics---gauge, spinor, twistor and other---are not postulated, but arise as a consequence of the fundamental relation \eqref{kass}.
It is especially important that equations~\eqref{kass} have a class of solutions corresponding to the light-like (shearfree) congruence generated by a point particle moving along some world lines. 
It can be considered as the Unique World Line (UWL) and naturally generates a whole set of identical point particles---duplicons, in the spirit of the ``single-electron Universe'' by Wheeler--Feynman. 
For an inertial observer and a polynomially parametrized UWL, the collective dynamics of duplicons is always conservative. 
Thus, one encounters a quite new approach to the natural description of systems of many particles and to fixing the true type of interparticle interactions.

Moreover, with the flow of the observer's time, against the background of general runaway, there are effects of pair formation followed by clustering. 
This makes it possible to approach the solution of another fundamental problem, the mechanism of formation of more and more complex material structure against the background of spread out (mutual recession) of their components.

In the future, it is planned to study a more general (and natural from the point of view of AD) case of collective dynamics of string-like singularities---caustics. 
Another prospect is to pursue the consideration of dynamics in the complete complexified space--time $\C \bf M$, the vector space of the algebra $\mathbb B$, and try to find out the meaning of the four extra dimensions.

\section*{Funding}
One of the authors (V.K.) received support from Project no. FSSF-2003-0003.

\end{document}